\documentclass[12pt,preprint]{aastex}

\newcommand{\rv}{{\bf r}}

\newcommand{\SigT}{\sigma_{\mathrm{LN}, \, T}}
\newcommand{\Sign}{\sigma_{\mathrm{LN}, \, n}}
\newcommand{\A}{\langle}
\newcommand{\E}{\rangle}

\newcommand{\Tew}{T_{\mathrm{ew}}}
\newcommand{\Tspec}{T_{\mathrm{spec}}}
\newcommand{\Tcl}{T_{\mathrm{cl}}}
\newcommand{\Dn}{\delta_n}
\newcommand{\DT}{\delta_T}
\newcommand{\rfit}{r_\sperp}
\newcommand{\rfitt}{r_{\mathrm{c,fit,3D}}}
\newcommand{\rfitsx}{r_{\mathrm{c,fit,Sx}}}
\newcommand{\Sx}{S_{\mathrm{x}}}
\newcommand{\betat}{\beta_{\mathrm{fit,3D}}}

\newcommand{\rvir}{r_{\mathrm{vir}}}
\newcommand{\dA}{d_{\mathrm{A}}}
\newcommand{\dAest}{d_{\mathrm{A,est}}}
\newcommand{\rc}{r_{\mathrm{c}}}

\newcommand{\Lambdax}{\Lambda_{\rm x} }

\newcommand{\thetac}{\theta_\mathrm{c}}
\newcommand{\thetacfit}{\theta_\mathrm{c,fit}}
\newcommand{\fH}{f_\H}

\newcommand{\T}{{\scriptscriptstyle T}}
\renewcommand{\H}{{\scriptscriptstyle H}}
\newcommand{\rcisobeta}{{r_{{\rm c,iso}\beta}}}
\newcommand{\rcpolybeta}{r_{{\rm c,poly}\beta}}
\newcommand{\rcfit}{{r_{{\rm c,fit}}}}

\newcommand{\rvd}{\tilde{r}}

\newcommand{\av}{{\bf a}}
\newcommand{\phia}{\phi_{\av}}
\newcommand{\thetaa}{\theta_{\av}}
\newcommand{\thetar}{\theta_{\rv}}

\newcommand{\Mr}{{\it M}_{\av} (\zeta)}
\newcommand{\xa}{x_\av}
\newcommand{\ya}{y_\av}
\newcommand{\za}{z_\av}

\newcommand{\sperp}{{\scriptscriptstyle\perp}}
\newcommand{\spara}{{\scriptscriptstyle\parallel}}

\slugcomment{}
\shortauthors{Kawahara et al.}
\shorttitle{Systematics in the Hubble Constant from SZE}
\begin{document}
%
\title{Systematic Errors in the Hubble Constant Measurement \\ 
from the Sunyaev-Zel'dovich effect}
\author{
 Hajime Kawahara\altaffilmark{1}, 
 Tetsu Kitayama\altaffilmark{2}, 
 Shin Sasaki\altaffilmark{3}, and
 Yasushi Suto\altaffilmark{1,4}} 
\altaffiltext{1}{Department of Physics, The University of Tokyo, 
Tokyo 113-0033, Japan}
\altaffiltext{2}{Department of Physics, Toho University,  Funabashi,
  Chiba 274-8510, Japan}
\altaffiltext{3}{Department of Physics, Tokyo Metropolitan University,
  Hachioji, Tokyo 192-0397, Japan}
\altaffiltext{4}{Research Center for the Early Universe, Graduate School
of Sciences, The University of Tokyo, Tokyo 113-0033, Japan}
\email{kawahara@utap.phys.s.u-tokyo.ac.jp}
\begin{abstract}
The Hubble constant estimated from the combined analysis of the
Sunyaev-Zel'dovich effect and X-ray observations of galaxy clusters is
systematically lower than those from other methods by 10-15 percent. We
examine the origin of the systematic underestimate using an analytic
model of the intracluster medium (ICM), and compare the prediction with
idealistic triaxial models and with clusters extracted from cosmological
hydrodynamic simulations.  We identify three important sources for the
systematic errors; density and temperature inhomogeneities in the ICM,
departures from isothermality, and asphericity.  In particular, the
combination of the first two leads to the systematic underestimate of
the ICM spectroscopic temperature relative to its emission-weighed one.
We find that these three systematics well reproduce both the observed
bias and the intrinsic dispersions of the Hubble constant estimated from
the Sunyaev-Zel'dovich effect.
\end{abstract}

\section{Introduction \label{sec:intro}}

Galaxy clusters constitute an important cosmological probe, in
particular in determining the Hubble constant $H_0$ through the combined
analysis of the Sunyaev-Zel'dovich effect (SZE) \citep{sunyaev72} and
X-ray observations.  Recent high-resolution X-ray and radio observations
enable one to construct a statistical sample of clusters for the $H_0$
measurement.  \citet{carlstrom02} compiled the previous results of 38
distance determination to 26 different galaxy clusters, and obtained
$H_0=60 \pm 3 \, \mathrm{km \, s^{-1} \, Mpc^{-1}}$ \citep[but see
Bonamente et al. 2006]{reese02,uzan04}. Despite its relatively large
individual errors, the mean value of $H_0$ estimated from SZE and X-ray
appears systematically lower than those estimated with other methods:
e.g. $H_0=72\pm8 \, \mathrm{km \, s^{-1} \, Mpc^{-1}}$ from the distance
to Cepheids \citep{freedman01}, and $H_0=73 \pm3 \, \mathrm{km \, s^{-1}
\, Mpc^{-1}}$ from the cosmic microwave background anisotropy
\citep{spergel07}.

Possible systematic errors in the $H_0$ measurement from the SZE have
been extensively studied by several authors 
\citep{inagaki95,kobayashi96,yoshikawa98,hughes98,birkinshaw99,wang07};
they have addressed a number of physical sources of possible biases
including the finite extension, clumpiness, asphericity, and
non-isothermality of the intracluster medium (ICM).  
Nevertheless they were not able to identify any systematic error that
affects the estimate of $H_0$ by 10-15 percent.  Therefore it has been
generally believed that the reliability of $H_0$ from the SZE is
dominated by the statistics.  Given that, the 10-15 percent
underestimate bias mentioned above, if real, needs to be explained in
terms of additional ICM physics beyond the simple models used in
previous studies.

 Recently, \citet{mazzotta04} have pointed out that the spectroscopic
temperature, $\Tspec$, is systematically lower than the
emission-weighted temperature, $\Tew$. \citet[hereafter Paper
I]{kawahara07} investigated the origin of the discrepancy and found that
both the fluctuation of density and temperature and non-isothermality
cause the difference between $\Tspec$ and $\Tew$. They also found that
the probability density functions (PDF) of both density and
temperature are well approximated by the log-normal function.

The aim of the paper is to revisit the origin of the bias based on an
observable quantity, $\Tspec$ and the log-normal description for
fluctuations of ICM. The paper is organized as follows. In \S 2, we
briefly review the conventional method to estimate $H_0$ from the
spherical isothermal $\beta$ modeling of galaxy clusters. Then we
describe several possible sources of the systematic bias based on the
log-normal description of the fluctuations and the spectroscopic
temperature. We propose an analytical model for the bias in \S 3.
Non-spherical effects are considered in \S 4 on the basis of triaxial
model clusters which include the log-normal fluctuation and the
temperature profile. Section 5 explores the validity of our analytic
model for the systematic bias using clusters extracted from cosmological
hydrodynamic simulations. Finally, we summarize our conclusions in \S 6.
Appendix A describes the semi-analytic distribution function of the bias
of the estimated Hubble constant due to the asphericity of clusters.

\section{Estimating $H_0$ from the SZE in the spherical 
isothermal $\beta$ model \label{sec:isobeta}}

A conventional estimate of $H_0$ from the SZE is based on the
assumptions that the gas temperature is isothermal, $T(r)= \Tcl$ (=
const.), and that the gas density follows the spherical $\beta$ model:
\begin{equation}
n(r) = n_0 \,\left[  1 + \left(\frac{r}{\rc}\right)^2  \right]^{-3 \beta/2},
\label{eq:beta-model}
\end{equation}
where $n_0$ is the central density, $r_c$ is the core radius, and
$\beta$ is the index characterizing the density profile.  These
approximations are insufficient to model the full complexity of real
galaxy clusters. It has been (implicitly) assumed that the average over
a number of clusters should significantly reduce the resulting error in
the estimate of $H_0$. While we will quantitatively argue below that
this is not the case, we summarize here the commonly adopted estimator
for $H_0$ in the spherical isothermal $\beta$ model
\citep{inagaki95,kobayashi96}.

In this idealistic model, the X-ray surface brightness and $y$-parameter
of SZE at an angle $\theta$ from the center of cluster are given by
\begin{eqnarray}
\label{eq:sx}
\Sx(\theta) &=&  \frac{\Lambdax(\Tcl) n_0^2 \rc G(\beta)}{4 \pi (1+z)^4}  
\left[1+ \left(\frac{\theta}{\thetac}\right)^2 
\right]^{-3 \beta + \frac{1}{2}}  \\
\label{eq:y}
y(\theta) &=& \frac{n_0 \sigma_\T \, k \Tcl \, \rc  G(\beta/2)}{m_e c^2}  
\left[1+ \left(\frac{\theta}{\thetac}\right)^2 
\right]^{-\frac{3}{2} \beta + \frac{1}{2}} ,
\end{eqnarray}
where $m_e$ is the electron mass, $k$ is the Boltzmann constant, $c$ is
the speed of light, $\sigma_\T$ is the Thomson cross section, $\Lambdax
(T)$ is the cooling function,  $z$ is
the redshift of the cluster, and we define
\begin{eqnarray}
 G(\beta) \equiv \sqrt{\pi} \, 
\frac{\Gamma(3\beta - 1/2)}{\Gamma(3 \beta)} 
\end{eqnarray}
with $\Gamma(x)$ being the gamma function.

Combining equations (\ref{eq:sx}) and (\ref{eq:y}), one can eliminate
$n_0$ and estimate the core radius as
\begin{eqnarray}
\label{eq:rcisobeta}
\rcisobeta (\Tcl) = 
\frac{y(0)^2}{\Sx(0)} \frac{m_e^2 c^4 \Lambdax(\Tcl)}{4 \pi
  (\sigma_\mathrm{T} k \Tcl)^2
 (1 + z)^4} \frac{G (\beta)}{[G (\beta/2)]^2} ,
\end{eqnarray} 
where $\Sx(0)$ and $y(0)$ denote the values at $\theta=0$, the
line-of-sight through the center of the galaxy cluster. Note that the
right-hand-side of equation (\ref{eq:rcisobeta}) is written entirely in
terms of observable quantities.

Equation (\ref{eq:rcisobeta}) corresponds to the estimate of the core
radius along the line-of-sight. If the cluster is spherically symmetric,
it should be equal to the core radius in the plane of the sky.
With the assumption, the measured angular core radius, $\thetacfit$, is
related to the physical core radius simply by
\begin{eqnarray}
\label{eq:rcfit}
r_{\rm c, fit} = \thetacfit \dA(z) 
 \end{eqnarray}
with $\dA(z)$ being the angular diameter distance of the cluster at $z$.
Equations (\ref{eq:rcisobeta}) and (\ref{eq:rcfit}) may be combined to
estimate the angular diameter distance to the cluster \citep{silk78}:
\begin{eqnarray}
\label{eq:daest}
\dAest(z) \equiv \frac{\rcisobeta}{\thetacfit} .
 \end{eqnarray}
If one obtains $\dAest(z)$ for a number of clusters at different
redshifts, one can estimate cosmological parameters by fitting to the
angular diameter distance vs. redshift relation, $\dA(z)$.  In what
follows, however, we consider the above methodology for the purpose of
estimating $H_0$. Thus following \citet{inagaki95}, we introduce the
ratio of the estimated to the true value of $H_0$:
\begin{eqnarray}
\label{eq:def-fh}
 \fH &=& \frac{\dA}{\dAest}= \frac{H_{\rm 0, est}}{H_{\rm 0, true}}
= \frac{r_\sperp}{r_\spara}.
\end{eqnarray}
Equations (\ref{eq:rcisobeta}) and (\ref{eq:rcfit})  provide commonly
used estimators for the radius of clusters along and perpendicular to
the line-of-sight, $r_\spara$, $r_\sperp$, respectively, but they are
model-dependent and ill-defined  for generic non-spherical
clusters. We will come back to this issue below (\S 4 and 5).  Note that
$\fH > 1$ ($<1$) corresponds to over(under)-estimate the true $H_0$.

Given the approximations underlying the spherical isothermal $\beta$
model, it is not surprising that $\fH$ for a individual cluster deviates
from unity. A more relevant question is whether the average over a
number of clusters, $\langle \fH \rangle$, is still systematically
larger or smaller than unity. If such systematic errors exist, can we
correct for them by identifying their physical origin ? This is what we
address in the present paper.

In fact there are several previous attempts toward the same goal, mainly
utilizing numerically simulated galaxy clusters
\citep{inagaki95,yoshikawa98,sulkanen99}. They concluded that departure
from the sphericity and the isothermality of clusters results in $\fH
\not= 1$, but after averaging over a sample the systematic errors are
relatively small, $|\langle \fH \rangle -1| \approx 5$\%.  Our analysis
below is different from the previous ones in adopting the spectroscopic
temperature, $\Tspec$, for $\Tcl$.  Indeed $\Tcl$ is a somewhat
ambiguous quantify for actual clusters (not isothermal).  It has been
common in this field to assume that the emission-weighted temperature:
\begin{equation}
\Tew \equiv \frac{\int n^2 \Lambdax (T) T dV}{\int n^2 \Lambdax (T) dV},
\label{eq:def_tew}
\end{equation}
is approximately equal to $\Tspec$ (the above integration is carried out
over the entire cluster volume). Thus the previous conclusion is
entirely based on the assumption that $\Tcl=\Tew$. Recently, however,
\citet{mazzotta04} and \citet{rasia05} pointed out that $\Tspec$,
estimated by fitting the thermal continuum and the emission lines of the
X-ray spectrum, is systematically lower than $\Tew$. Furthermore in
Paper I we found that the difference between $\Tew$ and $\Tspec$ could
be explained through an analytic model of the temperature profile and
inhomogeneities in the ICM.  We will evaluate
$\fH$ applying the model and then comparing the numerical simulations in
the subsequent sections.

\section{Analytic modeling of systematic errors of $H_0$ 
for spherical clusters
\label{sec:analytic}}

Identifying possible systematic errors in the estimate of $H_0$ for
realistic clusters is inevitably complicated. In order to address the
problem as analytically as possible, we consider spherical clusters that
follow a density profile of equation (\ref{eq:beta-model}) and a
polytropic temperature profile but with log-normal density and
temperature fluctuations. While the approach in this section is not
entirely generic, it is useful in understanding the physical origin of
systematic errors. The present analytic modeling will be tested against
numerically generated triaxial cluster samples in \S
\ref{sec:numerical}, and against those from cosmological hydrodynamic
simulations in \S \ref{sec:simulation}.

Our task here is to derive analytic expressions for more general
cases, which correspond to equations (\ref{eq:sx}) to
(\ref{eq:rcisobeta}) in the case of the isothermal $\beta$ model.  Let
us consider first the effect of inhomogeneities in ICM.  The X-ray
surface brightness at the center of the cluster is written as an
integral over the line-of-sight:
\begin{eqnarray}
\label{eq:sx0-1los}
 \Sx(0) = \frac{1}{4 \pi (1+z)^4} \int n(\rv)^2
  \Lambdax(T(\rv))d r .
\end{eqnarray}
Paper I found that the fluctuation fields defined as $\Dn \equiv
n(\rv)/n(r)$ and $\DT \equiv T(\rv)/T(r)$ are approximately independent
and follow the $r$-{\it independent} log-normal PDF, $P_{\rm
LN}(\Dn;\Sign)$ and $P_{\rm LN}(\Dn;\SigT)$, where $\Sign$ and $\SigT$
denote the standard deviations of the density and temperature
logarithms. The average of equation (\ref{eq:sx0-1los}) over many
independent line-of-sights can then be computed by integrating over the
log-normal PDFs. If we further assume that the cooling function,
 $\Lambdax (T)$,  is dominated by thermal
bremsstrahlung (bolometric), $\Lambda_{\rm bremss} (T) \propto
\sqrt{T}$, we can rewrite equation (\ref{eq:sx0-1los}) as
\begin{eqnarray}
 \label{eq:df}
 \Sx(0) &=& \frac{1}{4 \pi (1+z)^4} \int \Dn^2 \DT^{1/2} 
 P_{\rm LN}(\Dn) P_{\rm LN}(\DT)  d\Dn d\DT 
  \int n(r)^2   \Lambda_{\rm bremss}(T(r))  d r \cr
 &=& \frac{\exp{(\Sign^2 - \SigT^2/8)}}{4 \pi (1+z)^4} 
  \int n(r)^2   \Lambda_{\rm bremss}(T(r))  d r.
\end{eqnarray}
On the contrary, their fluctuations do not affect $y(0)$ because the
integrand of the $y$-parameter is a linear function of both temperature
and density. Thus the inhomogeneity effect is well described by the
factor:
\begin{eqnarray}
\label{eq:chisigma}
\chi_{\sigma} \equiv \exp{(\Sign^2-\SigT^2/8)}.  
\end{eqnarray}

The polytropic temperature profile is expressed as
\begin{equation}
T(r) = T_0 \, [n(r)/n_0]^{\gamma - 1},
\label{eq:polytropic}
\end{equation}
where $T_0$ is the central temperature (at $r=0$), and $\gamma$ is the
polytropic index. Then we obtain
\begin{eqnarray}
\label{eq:sx0_poly}
 \Sx(0) &=&  \chi_{\sigma} \frac{1}{4 \pi (1+z)^4} 
  \int n(r)^2   \Lambda_{\rm bremss}(T(r))  d r \cr
&=&  \chi_{\sigma} \frac{\Lambda_{\rm bremss}(T_0) n_0^2 
\rc G(\beta (\gamma+3)/4)}{4 \pi (1+z)^4} , 
\end{eqnarray}
and
\begin{eqnarray}
\label{eq:y0_poly}
y(0) &=& \frac{n_0 \sigma_{\rm T} k T_0 \rc  G(\beta \gamma/2)}{m_e c^2}, 
\end{eqnarray}
respectively.  Therefore, the core radius in this model is written as
\begin{eqnarray}
r_{\rm c, poly LN} &=&  \chi_{\sigma} \frac{y(0)^2}{\Sx(0)} 
\frac{m_e^2 c^4 \Lambda_{\rm bremss}(T_0)}
{4 \pi (\sigma_\mathrm{T} k T_0)^2 (1 + z)^4} 
\frac{G (\beta(\gamma+3)/4)}{[G(\beta \gamma/2)]^2} .
\label{eq:rcp}
\end{eqnarray}

If one attempts to fit the X-ray surface brightness profile under the
assumption of the isothermal $\beta$ model, the fitted value of the
$\beta$ parameter should be 
\begin{eqnarray}
\label{eq:betafit}
\beta_{\rm fit} = \frac{\beta (\gamma +3)}{4} ,
\end{eqnarray} 
since $\Lambda_{\rm bremss}[T(r)] n(r)^2 \propto T(r)^{1/2} n(r)^2
\propto [n(r)^{(\gamma +3)/4}]^2$. In addition, the fitted temperature
should be equal to the spectroscopic temperature $T_{\rm spec}$.  Thus
the estimated core radius is given by equation (\ref{eq:rcisobeta}):
\begin{eqnarray}
\label{eq:rcisobeta-polyLN}
\rcisobeta (\Tspec) = \frac{y(0)^2}{\Sx(0)} 
\frac{m_e^2 c^4 \Lambda_{\rm bremss}(\Tspec)}
{4 \pi  (\sigma_\mathrm{T} k \Tspec)^2 (1 + z)^4} 
\frac{G (\beta_{\rm fit})}{[G (\beta_{\rm fit}/2)]^2} .
\end{eqnarray} 
Therefore the systematic bias in the estimate of the Hubble constant in
this particular model should be
\begin{eqnarray}
\label{eq:fH_polyLNisobeta1}
f_{\rm H, polyLN|iso\beta} 
&=& \frac{r_{\rm c, poly LN}}{\rcisobeta (\Tspec)} 
= \chi_{\sigma}
\frac{\Lambda_{\rm bremss}(T_0)/T_0^2}{\Lambda_{\rm bremss}(\Tspec)/\Tspec^2}
\frac{G (\beta(\gamma+3)/4)}{[G(\beta \gamma/2)]^2} 
\frac{[G (\beta_{\rm fit}/2)]^2}{G (\beta_{\rm fit})} \cr
&=& \chi_{\sigma}
\frac{\Lambda_{\rm bremss}(T_0)/T_0^2}{\Lambda_{\rm bremss}(\Tspec)/\Tspec^2}
\left[\frac{G(\beta(\gamma+3)/8)}{G(\beta \gamma/2)}\right]^2 
\equiv \chi_{\sigma} \chi_{\rm T}(\Tspec),
\end{eqnarray}
where we define $\chi_{\rm T}$ that expresses the effect of the
temperature structure in the ICM. 

It may be more instructive to rewrite equation
(\ref{eq:fH_polyLNisobeta1}) as
\begin{eqnarray}
\label{eq:fH_polyLNisobeta2}
f_{\rm H, polyLN|iso\beta} 
= \chi_{\sigma} ~ \chi_{\rm T}(\Tew) ~
\frac{\chi_{\rm T}(\Tspec)}{\chi_{\rm T}(\Tew)} ,
\end{eqnarray}
since $T_{\rm cl}$ was often assumed to be equal to $\Tew$.  Equation
(\ref{eq:fH_polyLNisobeta2}) makes it clear that the systematic bias in
the estimate of $H_0$ results from three major effects; $\chi_\sigma$
due to inhomogeneities in the ICM, $\chi_{\rm T}(\Tew)$ representing the
temperature structure assuming that $T_{\rm cl}=\Tew$, and finally
$\chi_{\rm T}(\Tspec)/\chi_{\rm T}(\Tew)$ coming from the difference
from the spectroscopic and the emission-weighted temperatures of the
ICM.

Those three factors can be expressed in an approximate but analytic
fashion as follows. If we adopt the log-normal PDF for the density and
temperature inhomogeneities in the ICM, $\chi_{\sigma}
=\exp(\Sign^2-\SigT^2/8)$ (eq.[\ref {eq:chisigma}]). As shown in Paper
I, cosmological hydrodynamic simulations indicate that $\Sign \approx
0.2$--$0.5$ and $\SigT \approx 0.2$--$0.3$. Thus $\chi_{\sigma} \approx
1.04$--$1.3$. The second factor can be estimated by using the analytical
relation of $T_0$ and $\Tew$ in the current model:
\begin{eqnarray}
\Tew/T_0 &=& \exp{(\SigT^2/2)} \,  J(\beta,\gamma,r_c/\rvir) ,
\end{eqnarray}
where we assume that the cluster has a finite extension and $n(r)=0$ for
the radius $r$ beyond the virial radius of the cluster, $\rvir$, and we
define
\begin{eqnarray}
 J(\beta,\gamma,r_c/\rvir) &\equiv& 
\frac{\,_2 F_1 (3/2,3 \beta [1 + 3(\gamma-1)/4 ]
 ; 5/2 ; -\rvir^2/r_c^2 )}{\,_2 F_1 (3/2,3 \beta [1
 + (\gamma-1)/4] ; 5/2 ; -\rvir^2/r_c^2 )}, 
\label{eq:tewp_poly}
\end{eqnarray}
with $\,_2 F_1 ( \alpha,\beta; \gamma; \zeta)$ being the hyper-geometric
function (see \S 3 of Paper I).  Just for simplicity, we neglect the
term, $\exp{(\SigT^2/2)}$, that represents the temperature inhomogeneity
because it is relatively small for $\SigT \approx 0.2-0.3$.  If we
further adopt  $\Lambdax (T) = \Lambda_{\rm bremss}(T)\propto \sqrt{T}$  then $\chi_{\rm
T}(\Tew)$ reduces to
\begin{eqnarray}
\label{eq:chiT}
\chi_{\rm T}(\Tew)
&\approx&  \left(\frac{\Tew}{T_0}\right)^{1.5} 
\frac{[G(\beta (\gamma +3)/8)]^2}{[G(\beta \gamma/2)]^2} \cr
&=&  J(\beta,\gamma,r_c/\rvir)^{1.5} 
\frac{[G(\beta (\gamma+3)/8)]^2}{[G(\beta \gamma/2)]^2}.
\end{eqnarray}

\begin{figure}[tbhp]
{\plotone{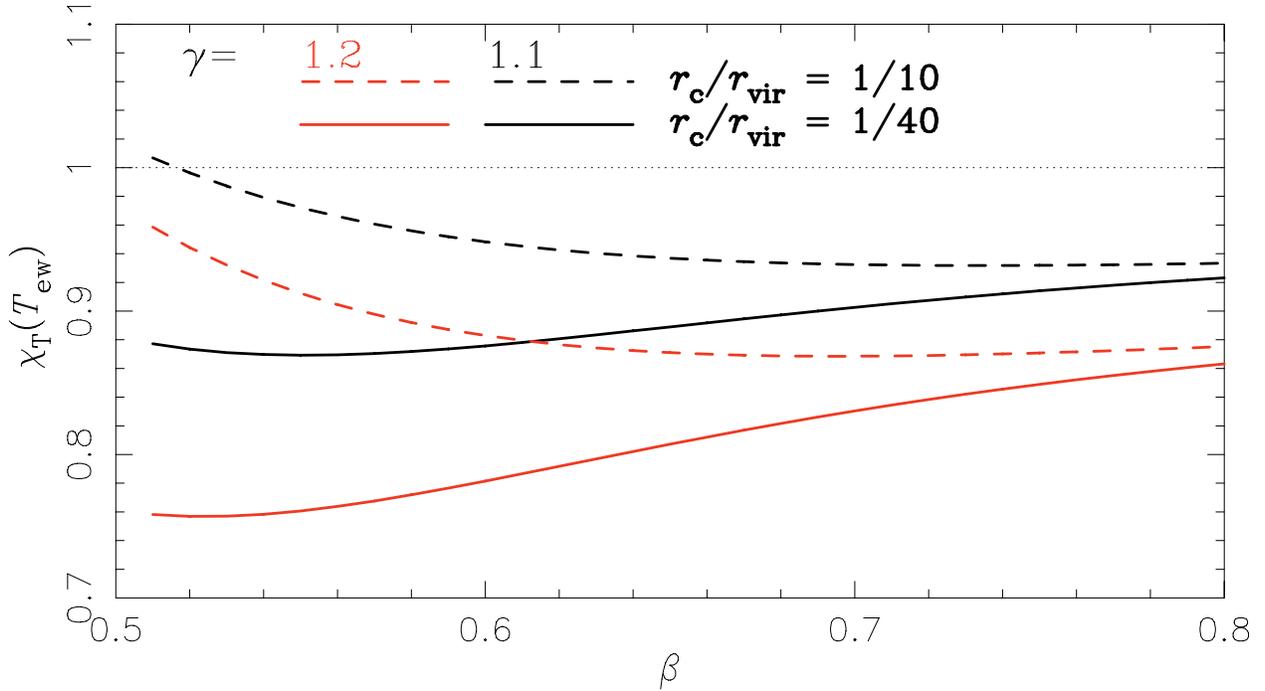}} 
\caption{The bias of $H_0$ due to the temperature profile
(eq.\ref{eq:chiT}) as a function of $\beta$.  Dashed and solid curves
correspond to the cases that core radius is 10 percent and 2.5 percent
of the virial radius, respectively. Red and black colors indicate the
polytropic index $\gamma=1.2$ and $1.1$, respectively.  } \label{fig:p}
\end{figure}

The result is plotted in Figure \ref{fig:p} for typical values of the
parameters, and indicates that $\chi_{\rm T}(\Tew)$ ranges from 0.8 to
1.0 for $\beta = 0.5-0.8$ and $\gamma=1.1-1.2$.

Similarly the third factor can be approximated as
\begin{eqnarray}
\label{eq:spec-ew}
\chi_{\rm spec-ew} \equiv
\frac{\chi_{\rm T}(\Tspec)}{\chi_{\rm T}(\Tew)} 
= \frac{\Tspec^2}{\Tew^2} \, 
\frac{\Lambda_{\rm bremss}(\Tew)}{\Lambda_{\rm bremss}(\Tspec)}
\approx \left(\frac{\Tspec}{\Tew}\right)^{1.5}.
\end{eqnarray}
Several studies confirmed the systematic underestimate of the
spectroscopic temperature relative to the emission-weighted temperature,
$\Tspec/\Tew= 0.8 -0.9$ from cosmological hydrodynamic simulations
\citep[paper I]{rasia05,kay06}.  If $\Tspec/\Tew=$ 0.8 (0.9),
for instance, $\chi_{\rm spec-ew}$ amounts to 0.7 (0.85).

\section{Numerical modeling of systematic errors of $H_0$ 
for inhomogeneous and triaxial clusters \label{sec:numerical}}

So far we have only considered spherical clusters. Asphericity is
definitely another important source of error in the estimate of
$H_0$. The errors are expected to be significantly reduced by averaging
over a statistical sample of clusters randomly oriented with respect to
our line-of-sight. Nevertheless, if clusters preferentially take either
 prolate or oblate shapes, for instance, the residual errors may not be
entirely negligible. This is why we address the effect of asphericity on
the basis of the triaxial approximation for the cluster ICM
\citep{hughes98,jingsuto02,lee03,lee04}.
 
To investigate quantitatively the combined effects of gas inhomogeneity
and asphericity, we numerically create three sets of cluster samples and
perform mock observations of Monte-Carlo realizations. The first one is
spherical, but include random gas density and temperature fluctuations
according to the log-normal distribution. The second one is triaxial
without the fluctuations. Both the first and the second samples assume
isothermality ($\gamma=1$). The third one corresponds to the model
described in \S 3 except for added asphericity; the log-normal
fluctuations, the polytropic temperature structure, and the triaxiality
are included.  We call them {\it model clusters} in order to distinguish
them from {\it simulated clusters} extracted from cosmological
hydrodynamic simulations (\S 5).

Each model cluster is constructed on $(512)^3$ grid points within 6
$\mathrm{h^{-1} Mpc}$ cubic region around the center. We first create
spherically symmetric clusters with the gas density profile following
equation (\ref{eq:beta-model}) with $\beta = 0.65$, $r_{\rm c} = 100
\mathrm{h^{-1} kpc}$, and $n_0 = 10^{-2} \mathrm{cm^{-3}}$. The gas is
fiducially isothermal with $T=5$ keV, while we also consider the case of
polytropic temperature profile with $T_0=7$ keV and $\gamma=1.2$. We
then add random fluctuations of gas density and temperature according to
the $r$-independent log-normal distributions.  The X-ray emissivity is
computed with SPEX version 2.0 assuming collisional ionization
equilibrium, the energy range of $0.5-10.0$ keV and a constant
metallicity $0.3 Z_\odot$.  Triaxial model clusters are constructed
simply by stretching spherical clusters along the three axis directions
by a factor of $\lambda_a$, $\lambda_b$, and $\lambda_c$, respectively.

In mock observations,  we extract the quantities necessary
to compute $\rcisobeta (\Tcl)$ and $\rcfit$ via equations
(\ref{eq:rcisobeta}) and (\ref{eq:rcfit}) in the following manner. We
first fit the projected profiles of $\Sx(\theta)$ with a functional form
$\Sx(0) [1+(r/\rfitsx)^2]^{-3 \beta_{\rm fit} +1/2}$ from $r =0$ to $r =
\mathrm{1 h^{-1} Mpc}$ over 1024 random LOSs toward each cluster. For
each LOS, we also compute $y(0)$ and, unless otherwise stated, use it
directly in our analysis.  We will discuss other choices of obtaining
$y(0)$ in \S 5.3.   As will be described later, the gas
temperature $T_{\rm cl}$ is obtained by either fitting the mock X-ray
spectra or simply using the input temperature, depending on the purpose
of the analysis. We use the template of the spectral energy distribution
computed using SPEX 2.0 assuming collisional ionization equilibrium, the
energy range of $0.5-10.0 \mathrm{keV}$ and a constant metallicity $0.3
Z_\odot$. Assuming that $r_\spara=\rcisobeta (\Tcl)$ and
$r_\sperp=\rfitsx$, we calculate $\fH$ for each LOS.

To quantify the bias due to the projection effect, we also compute the
volume-averaged radial profile of the gas density, directly from the
grid data within the radius $\mathrm{1 h^{-1} Mpc}$. By fitting the
profile to the $\beta$ model, we obtain the estimated core radius
$\rfitt$, which is independent of LOS.   We will compare the
values of $\fH$ using $\rfit=\rfitt$ and $\rfitsx$ in what follows.

\begin{figure}[htbp]
{\plotone{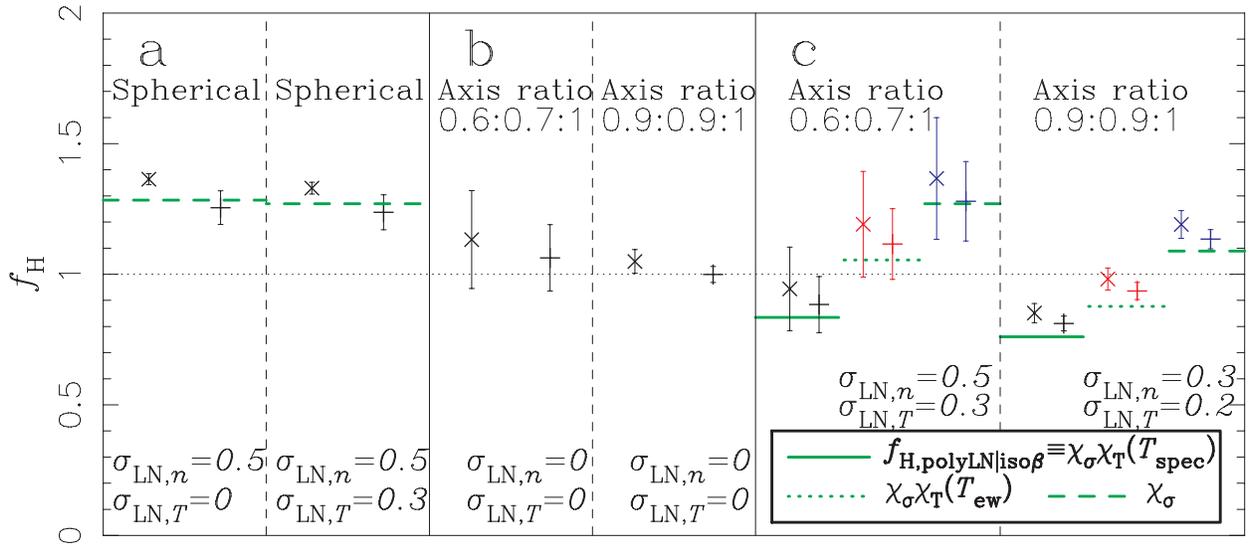}} \caption{The average and rms of $\fH$ of the model
clusters; (a) spherical clusters with gas inhomogeneities and no
temperature gradient, (b) ellipsoidal and isothermal clusters, and (c)
ellipsoidal clusters with temperature gradient and gas inhomogeneities.
Crosses and pluses denote $\fH$ adopting $r_\sperp=\rfitsx$ and
$r_\sperp=\rfitt$, respectively.  Thick horizontal lines indicate
analytical estimations for $\chi_\sigma$ (dashed), $\chi_\sigma
\chi_{\rm T}(T_{\rm ew})$ (dotted), and $\chi_\sigma \chi_{\rm T}(T_{\rm
spec})$ (solid). In panel (c), black symbols indicate $\fH$ using
$\rcisobeta(\Tcl=\Tspec)$, red symbols $\rcisobeta(\Tcl=\Tew)$, and blue
symbols $\rcpolybeta$, which correspond to the isothermal fit with
$T=T_{\rm spec}$, the isothermal fit with $T=T_{\rm ew}$, and the
polytropic fit, respectively (see the main text for
details). \label{fig:imo}}
\end{figure}

Figure \ref{fig:imo}a shows the mean and rms values of $\fH$ for
spherical clusters with no temperature gradient.  We consider two cases
for the log-normal density and temperature fluctuations with
$(\Sign,\SigT) = (0.5,0.0)$ and $(0.5,0.3)$. The latter set corresponds
to the typical value for the simulated cluster as Paper I reported.  To
present the bias produced solely by gas inhomogeneities, we here adopt
for $\Tcl$ the volume averaged temperature instead of fitting the mock
X-ray spectra. It is evident that the fluctuations in gas density yield
$\fH \sim 1.3$ (i.e.  overestimating of $H_0$ by $\sim 30\%$), while
those in gas temperature do not contribute significantly to the
bias. The mean value of $\fH$ is in good agreement with our analytical
expectation $\chi_\sigma$ (dashed horizontal lines).  The bias due to
the projection is $\sim 10\%$.

The bias produced by ellipsoidal shapes is displayed in Figure
\ref{fig:imo}b for the two sets of the axis ratio
($\lambda_a:\lambda_b:\lambda_c = 0.6:0.7:1$ and $0.9:0.9:1$). These
sets are the typical value of the simulated cluster. Again, to present
the bias solely from asphericity, the gas is assumed to be isothermal
without any fluctuations and we adopt $\Tcl = T_0$ in computing $\fH$.
The average bias is relatively small ($ \lesssim 15 \%$) and that due to the
projection is $\sim 3\%$. These results indicate that the bias due to
asphericity, after averaging over a statistical sample of clusters, is
smaller than that from gas inhomogeneity.

Figure \ref{fig:imo}c illustrates the bias in a more realistic case;
we create ellipsoidal clusters with the polytropic temperature profile
and fluctuations. Two sets of axis ratio ($\lambda_a:\lambda_b:\lambda_c
= 0.6:0.7:1$ and $0.9:0.9:1$) are chosen adopting $(\Sign,\SigT) =
(0.5,0.3)$ and $(0.3,0.2)$, respectively.  In this panel, we show the
values of $\fH$ based on the following three methods, so as to
understand clearly the physical origin of the overall bias in the $H_0$
estimation.

The first method (black symbols in Fig. \ref{fig:imo}c) corresponds to
the most conventional case of using the isothermal $\beta$-model and the
spectroscopic temperature $T_{\rm spec}$. To obtain $T_{\rm spec}$ we
fit the mock X-ray spectra from the central ($r< 1$ $\mathrm{h^{-1}
Mpc}$) region of each cluster using XSPEC assuming a single temperature
MEKAL model. We assume the perfect response, and ignore its effect on
the spectral temperature (Paper I).  Clearly, the value of $H_0$ is {\it
underestimated} by $\sim 10$-$20\%$.  This is in good agreement with our
analytical estimation for $f_{\rm H, polyLN|iso\beta}=\chi_\sigma
\chi_{\rm T} (T_{\rm spec})$ for a spherical cluster (solid horizontal lines). To
obtain $\chi_{\rm T}(\Tspec)$, we compute the volume-averaged profile of
the density and the temperature. These profile are fitted to equation
(\ref{eq:beta-model}) and equation (\ref{eq:polytropic}) taking $n_0$,
$\rfitt$, $\beta$, $T_0$ and $\gamma$ as free parameters. We use the
adopted values $(\Sign,\SigT) = (0.5,0.3)$ and $(0.3,0.2)$ to compute
$\chi_{\sigma}$.

The second method (red symbols in Fig. \ref{fig:imo}c) aims to mimic
previous numerical studies of the $H_0$ bias
\citep{inagaki95,yoshikawa98} and adopts the isothermal $\beta$-model
and the emission-weighted temperature $T_{\rm ew}$.  We obtain $\Tew$ by
directly summing up the temperature of each grid point from the central
($r< 1$ $\mathrm{h^{-1} Mpc}$) region. Also plotted for comparison is an
analytical estimate $\chi_\sigma \chi_{\rm T} (T_{\rm ew})$ (dotted
lines). The values of $\chi_{\sigma}$ and $\chi_{\rm T} (T_{\rm ew})$
are computed as described above.  In this case, $\chi_{\sigma}$ and
$\chi_{\rm T} (T_{\rm ew})$practically cancel each other, and $\fH$ is
close to unity, consistent with the previous findings of
\citet{inagaki95} and \citet{yoshikawa98}. This shows that the absence
of the bias in previous studies is simply an artifact of using $\Tew$,
which is systematically larger than $T_{\rm spec}$.

The third method (blue symbols in Fig. \ref{fig:imo}c) attempts to
eliminate the bias due to the temperature gradient by using the
polytropic profile to estimate the core radius \citep{ameglio06} :
\begin{eqnarray}
\rcpolybeta &=& \frac{y(0)^2}{\Sx(0)} 
\frac{m_e^2 c^4 \Lambdax(T_0)}{4 \pi (\sigma_\mathrm{T} k T_0)^2
 (1 + z)^4} \frac{G (\beta \gamma /4 + 3 \beta /4)}{[G(\beta
 \gamma/2)]^2}, 
\end{eqnarray}
where we adopt $T_0$ and $\gamma$ from fitting the volume-averaged
temperature profile of the model clusters. The value of $\beta$ is
obtained from equation (\ref{eq:betafit}) using $\beta_{\rm fit}$ and
$\gamma$.  The value of $\fH$ so obtained should represent the bias
arising from sources {\it other than} the spectral fitting and the
temperature gradient. Given the good agreement with the analytical
estimate for $\chi_\sigma$ (dashed lines), we conclude that the bias in
this case is dominated by the effect from gas inhomogeneities.

In summary, there are three major sources for the bias of $H_0$; the
spectral fitting, the temperature gradient, and local density
fluctuations. The former two leads to an {\it underestimate} while the
latter an {\it overestimate} of $H_0$. In every case studied here, the
bias due to asphericity is much smaller than the other three.

\section{Comparison with clusters from cosmological hydrodynamic
 simulations
\label{sec:simulation}}

\subsection{Cosmological hydrodynamic simulations}

\begin{figure}[tbp]
{\plotone{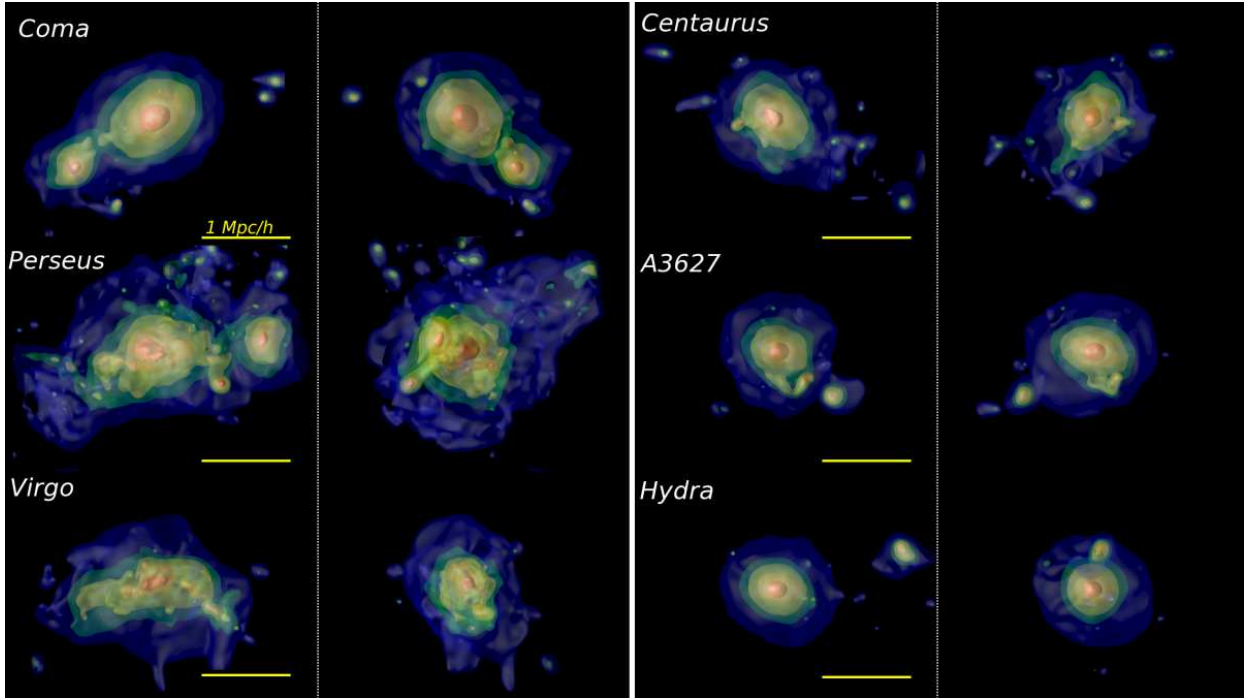}} \caption{Projected surface density maps of the six simulated
clusters. Five different ($n_e = 3 \times 10^{-3}, 1 \times 10^{-3}, 5
\times 10^{-4}, 3 \times 10^{-4}$, and $ 1 \times 10^{-4}$
[$\mathrm{cm^{-3}}$]) isodensity surfaces are indicated with different
colors (red, orange, yellow, green, and blue, respectively).  The left
panels of each cluster indicate the view from our galaxy. The right
panels are the projection of each simulated cluster as seen by a distant
observer located to the ``right'' of each panel on the left. The 
horizontal yellow lines indicate the physical size of 
$1 \, \mathrm{h^{-1} Mpc}$. \label{fig:ac}}
\end{figure}

We now compare the bias described in the previous section with simulated
clusters.  They are extracted from the Smoothing Particle Hydrodynamic
(SPH) simulation of the local universe performed by \citet{dolag05}
assuming $\Lambda $CDM universe with $\Omega_{0m}=0.3, \Omega_b=0.04,
\sigma_8=0.9$, and $h=0.7$. The numbers of dark matter and SPH particles
are $\sim 20$ million each within a high-resolution sphere of radius
$\sim 110 Mpc$, which is embedded in a periodic box $\sim 343$ Mpc on a
side that is filled with nearly 7 million low-resolution dark matter
particles. The simulation is designed to reproduce the
matter distribution of the local universe adopting the initial
conditions based on the IRAS galaxy distribution, smoothed over a scale
of $4.9 h^{-1} \mathrm{Mpc}$. We choose the six massive clusters
identified as Coma, Perseus, Virgo, Centaurus, A3627, and Hydra. Figure
\ref{fig:ac} shows projected surface density maps of these simulated
clusters.  The values of $\beta$ and $\gamma$ of these
clusters are listed in Table \ref{tab:axisratio}.  The
cubic region of 6 $h^{-1}$ Mpc around the center of each cluster is
extracted and divided into $512^3$ cells.  The density and temperature
of each mesh point are calculated from SPH particles using the B-spline
smoothing kernel.  A detailed description of this procedure is given in
Paper I.

We perform mock observations over 1024 LOSs for each simulated cluster
in a similar manner to \S \ref{sec:numerical} except for the following
points. First, we compute $\Tspec$ and $\Tew$ within the virial radius
instead of 1 $\mathrm{h^{-1} Mpc}$.  Second, we use the fitted value of
$\Sign$ and $\SigT$ in calculating $\chi_\sigma$ of the analytical
model.
\begin{table}[h]
  \caption{Properties of the Six Simulated Clusters.}
\label{tab:axisratio}
 \begin{center}
  \begin{tabular}{lcccccc}
   \hline\hline
   & $\betat$ $\,^*$ & $\gamma$ $\,^*$& $\lambda_a/\lambda_c$ & $\lambda_b/\lambda_c$ & $\A \rfitsx \E/\rfitt$\\
   \hline 
    Coma & 0.74 & 1.17 & 0.59 & 0.64 & 1.03 $\pm$ 0.14 \\ 
   Perseus & 0.64 & 1.09 & 0.49 & 0.61 & 1.04 $\pm$ 0.18 \\  
  Virgo & 0.60 & 1.15 & 0.44 & 0.61 & 1.16 $\pm$ 0.31 \\
Centaurus & 0.69 & 1.17 &  0.68 & 0.78 &  1.03 $\pm$ 0.13 \\
A3627 &  0.69 & 1.15 & 0.79 & 0.83 &  1.08 $\pm$ 0.06\\
 Hydra &  0.70 & 1.22 & 0.84 & 0.93 &  1.03 $\pm$ 0.05 \\
  \hline
{\hbox to 0pt{\parbox{200mm}{\footnotesize
$\ast$  The values of $\betat$ and $\gamma$ are slightly changed from that
   listed in Paper I \\
due to the improvement of the routine of fits. 
   }\hss}}
  \end{tabular}
 \end{center}
\end{table}

\subsection{Results}

\begin{figure}[htbp]
{\plotone{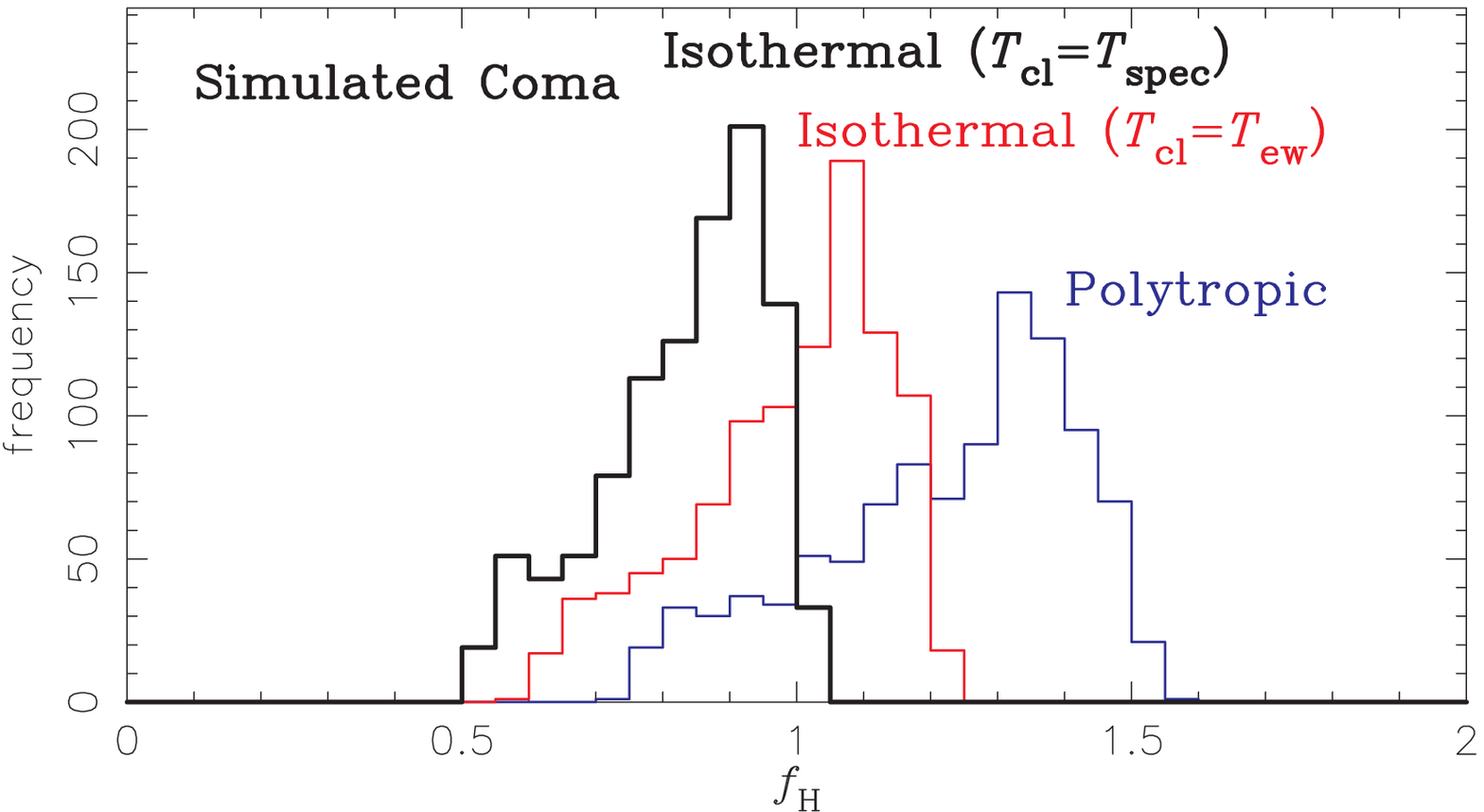}} 
\caption{Distribution of $\fH$ over 1024 LOSs for the simulated Coma
 cluster. Black, red and blue histograms indicate the results for the
 isothermal fit with $T=T_{\rm spec}$, the isothermal fit with $T=T_{\rm
 ew}$, and the polytropic fit, respectively.}  \label{fig:dist}
\end{figure}

Figure \ref{fig:dist} displays a set of histograms of $\fH$ for the
simulated Coma cluster. The same analysis is done for the other five
clusters.  Histograms in different colors correspond to the symbols of
the same color in Figure \ref{fig:imo}c, and indeed show similar trends
for each component of the bias.  Since the physical length of clusters
along the LOS is not symmetrically distributed around its mean, the
corresponding histograms of $\fH$ are skewed positively. In Appendix A,
we compute the distribution for the two extreme cases, the prolate and
the oblate ellipsoids, and find that they yield positively and
negatively skewed distributions, respectively.  Indeed this is
consistent with the fact that the simulated Coma is nearly prolate
(Table \ref{tab:axisratio}) .

\begin{figure}[htbp]
{\plotone{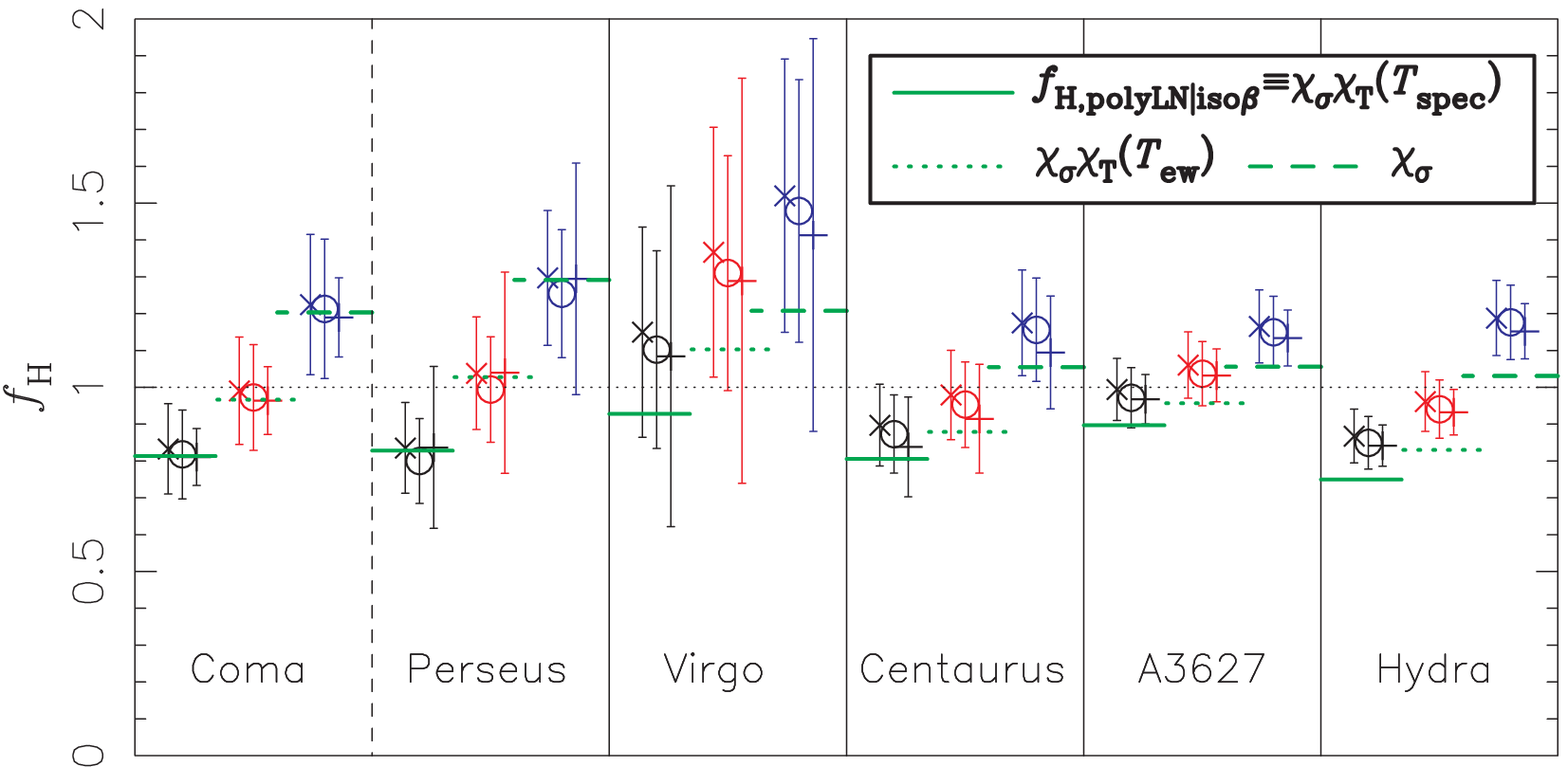}} 
\caption{The average and rms of $\fH$ for the six simulated clusters.
Black, red and blue symbols with error-bars indicate the results for the
isothermal fit with $T=T_{\rm spec}$, the isothermal fit with $T=T_{\rm
ew}$, and the polytropic fit, respectively.  Crosses and pluses denote
$\fH$ adopting $r_\sperp=\rfitsx$ and $r_\sperp=\rfitt$, respectively.
Open circles indicate $\fH$ adopting $r_\sperp=\rfitsx$ but assuming
that the clusters are not extended beyond the virial radius.  Thick
green horizontal lines indicate analytical estimations for $\chi_\sigma$
(dashed), $\chi_\sigma \chi_{\rm T}(T_{\rm ew})$ (dotted), and
$\chi_\sigma \chi_{\rm T}(T_{\rm spec})$ (solid). } \label{fig:avestd}
\end{figure}

The (simple arithmetic) mean, $\langle \fH \rangle$ is plotted in Figure
\ref{fig:avestd} for six simulated clusters. The quoted error bars
indicate 1$\sigma$ standard deviation from the mean.  Except for the
simulated Virgo cluster, $\langle \fH \rangle$ is below unity, i.e., $H_0$ is
{\it underestimated}.  It is remarkable that a simple analytical model
for systematic effects (solid, dotted and dashed horizontal lines)
described in \S \ref{sec:analytic} can reproduce the bias in the
simulated clusters.

We have made sure that the bias from other sources is minor; first, if a
cluster has a finite extension and is bounded within the virial radius,
the value of $\langle \fH \rangle$ becomes smaller by $\lesssim 5\%$ (open
circles in Fig.  \ref{fig:avestd}).  Second, we compute the axis ratio
($\lambda_a < \lambda_b < \lambda_c$) of each simulated cluster,
basically following the method of \citet{jingsuto02}, but using the gas
density not the dark matter density.  The isodensity surfaces
corresponding to the gas densities of $n_e = 3 \times 10^{-3}$, $1
\times 10^{-3}$, $5 \times 10^{-4}$, $3 \times 10^{-4}$, and $1 \times
10^{-4}$ [$\mathrm{cm^{-3}}$] are shown in Figure \ref{fig:ac}.  After
eliminating substructures, the axis ratio is calculated by diagonalizing
the inertial tensor of each surface. The averaged axis ratios of the
five different density regions (Table \ref{tab:axisratio}) are similar
to the typical value adopted in \S 4. Therefore, we conclude that the
spherical approximation itself is not a major source of the bias for the
simulated cluster. Finally, the bias due to the projection is small
(crosses and pluses in Fig.  \ref{fig:avestd}). We list in Table
\ref{tab:axisratio} the average values of $\rfitsx$ over 1024 LOSs
relative to $\rfitt$.  The ratio is unity within 10 \% (except for Virgo
that has a relatively large dispersion), and basically all
consistent with unity within the uncertainty.

 It is interesting to emphasize here that the shape of the
distribution of $\fH$ reflects the shape of clusters from the
perspective of measuring the three-dimensional shape of clusters
\citep[e.g.]{sereno06}.  If all clusters had the same shape, the
observation of one cluster toward {\it multiple directions } might
correspond to that of {\it multiple clusters} toward each LOS. Of
course, the real shapes vary from cluster to cluster. However, if
clusters tend to be prolate (oblate) preferentially, the distribution of
$H_{\rm 0,est}$ should be skewed positively (negatively) as shown in
appendix A. Therefore, independently of the knowledge of real value of
$H_{\rm 0,true}$, the statistical information about the shape
distribution may be obtained {\it in principle} by the distribution of
$H_{\rm 0,est}$.

\subsection{Comparison with previous studies}

The above results are consistent with the previous results of $\fH \sim
1$ with $\Tcl=\Tew$ \citep{inagaki95,yoshikawa98}.  On the other hand,
\citet{ameglio06} explored the bias of $\dA$ using the cosmological
hydrodynamic simulations, and reported that $H_0$ is {\it overestimated}
by more than a factor of two if one adopts the isothermal $\beta$
model. This is opposite to our conclusion here, and we found that this
should be ascribed to the sensitivity of $f_{\rm H}$ on the adopted
values of $y(0)$, i.e., $ f_{\rm H} \propto \dAest^{-1} \propto
y^{-2}(0)$ as explained below. 

\citet{ameglio06} obtained $y(0)$ by fitting the noise-less profile of
$y(\theta)$ up to $R_{500}$ fixing other $\beta$-model parameters from
the X-ray profile, while we use directly the projected values of $y(0)$
in the simulation data. The difference in these two methods is apparent
in Figure 5 (right panel) of \citet{ameglio06}; their fit (solid line)
was affected largely by the data points at large radii and yielded a
value of $y(0)$ smaller by $\sim 50\%$ than the actual data. This
enhances the value of $f_{\rm H}$ by more than a factor of two, and
indeed accounts for their apparently opposite conclusions. We have
checked that other differences between their analysis and ours (the use
of mass-weighted temperature for $\Tcl$ and the removal of the cluster
central region) do not affect the results significantly.

As far as the bias in the previous SZE observations is concerned, we
believe that our method is more relevant to what has been done with the
real data, because these observations were not capable of constraining
the radial profile of the y-parameter up to large radii with high S/N
\citep[see][for the currently highest angular-resolution observation of
the SZE]{komatsu01,kitayama04}. 

We have further made sure that the effects of the finite spatial
resolution of the observations and the central cooling region, which
were neglected in our preceding analysis, are minor; first, we have
evaluated $y(0)$ by fitting $y(\theta)$ within a radius of 100 $h^{-1}
\mathrm{kpc}$ and 200 $ h^{-1} \mathrm{kpc}$. These approximately
correspond to the typical angular resolution of the SZE observation
($\sim$ 1 minute) at $z=0.1$ and $z=0.3$, respectively. The values of
$\beta$ and $\theta_c$ are fixed from the X-ray profile.  For the six
simulated clusters, the resulting values of $f_{\rm H}$ differ from our
initial analysis (see Fig. \ref{fig:avestd}) by $-6\%$ to $+7\%$ ($+2\%$
on average) for $r< 100h^{-1} \mathrm{kpc}$, and by $-3\%$ to $+10\%$
($+5\%$ on average) for $r< 200h^{-1} \mathrm{kpc}$.

Second, we have also performed the fitting separately for
the X-ray and SZE profiles. The values of $\Sx$, $\beta$ and $\theta_c$
are evaluated by fitting $\Sx(\theta)$ with equation (\ref{eq:sx}),
while that of $y(0)$ is obtained by fitting $y(\theta)$ with equation
(\ref{eq:y}) independently of the X-ray profile. As a
result, the values of $f_{\rm H}$ differ from those of Figure
\ref{fig:avestd} by $-6\%$ and $+1\%$ ($-3\%$ on average). 

Note that observationally there are several different ways to evaluate
$y(0)$, $\Sx (0)$, $\beta$ and $\theta_c$. A conventional method is to
fit the X-ray imaging data $\Sx(\theta)$ first. Then the SZ image is
fitted to obtain $y(0)$ assuming the values of $\beta$ and $\theta_c$
from the X-ray data. Our analysis procedure adopted here follows the
conventional method.  While \citet{reese02} have determined $d_A$ from
the joint fit to the X-ray and SZE imaging data, the result is almost
equivalent to the conventional method since the X-ray imaging data have
a much higher S/N than the SZE data.

\section{Conclusions \label{sec:conclusions}}

We considered various possible systematic errors of $H_0$ from the
combined analysis of the Sunyaev-Zel'dovich effect and X-ray
observations.  In particular we addressed the validity and limitation of
the spherical isothermal $\beta$ model in estimating $H_0$, which has
been used widely as a reasonable approximation after averaging over a
number of clusters. We introduced the ratio of the estimated to the true
Hubble constant, $f_H$, to characterize the systematic errors.  We
constructed an analytic model for $f_H$, and identified three important
sources for the systematic errors; density and temperature
inhomogeneities in the ICM, the temperature profile, and departures from
sphericity. Except for the non-spherical effect, the most important
analytical expression that summarizes our conclusion is equation
(\ref{eq:fH_polyLNisobeta2}), or equivalently,
\begin{eqnarray}
\frac{H_{\rm 0, est}}{H_{\rm 0, true}}
= \chi_{\sigma} ~ \chi_{\rm T}(\Tew) ~\chi_{\rm spec-ew} .
\end{eqnarray}
In our analytic model discussed in \S 3, the inhomogeneity bias,
$\chi_{\sigma}$, the non-isothermality bias, $\chi_{\rm T}(\Tew)$, and
the temperature bias $\chi_{\rm spec-ew}$ are given by equations
(\ref{eq:chisigma}), (\ref{eq:chiT}), and (\ref{eq:spec-ew}),
respectively.

While the above model prediction is fairly general, the net value of
$f_{\rm H}$ sensitively depends on the degree of the inhomogeneity and
multi-phase temperature structure of real ICM. Our simulated cluster
sample implies that $\chi_{\sigma} \approx (1.1-1.3)$, $\chi_{\rm
T}(\Tew) \approx (0.8-1)$, $\chi_{\rm spec-ew} \approx (0.8-0.9)$, and
therefore $\langle f_{\rm H}\rangle \approx (0.8-0.9)$. Given the result of
\citet{reese02}, this is certainly indicative, but may need to be
interpreted with caution because the result is critically dependent on
the reliability of the adopted numerically simulated clusters as
representative samples of clusters observed in the real universe.
Exactly for this reason, we are attempting more direct (not statistical)
comparison of our model prediction against observed cluster samples,
which will be presented elsewhere hopefully in the near future (Reese et
al. in preparation).

\bigskip 
\bigskip 

We thank Noriko Yamasaki and Kazuhisa Mitsuda for useful discussions,
Klaus Dolag for providing a set of simulated cluster samples, and Erik
Reese for a careful reading of the manuscript.  We also thank
an anonymous referee for several constructive comments. 
The simulations were performed at the Data-Reservoir at the University
of Tokyo, and we thank Mary Inaba and Kei Hiraki for providing the
computational resources.  This work is supported by Grant-in-Aid for
Scientific research of Japanese Ministry of Education, Culture, Sports,
Science and Technology (Nos. 14102204, 15740157, 16340053, 18740112, and
18072002), and by JSPS (Japan Society for Promotion of Science)
Core-to-Core Program ``International Research Network for Dark Energy''.

\appendix

\section{Distribution of $\fH$ for prolate and oblate ellipsoids}

\begin{figure}[htbp]
{\plotone{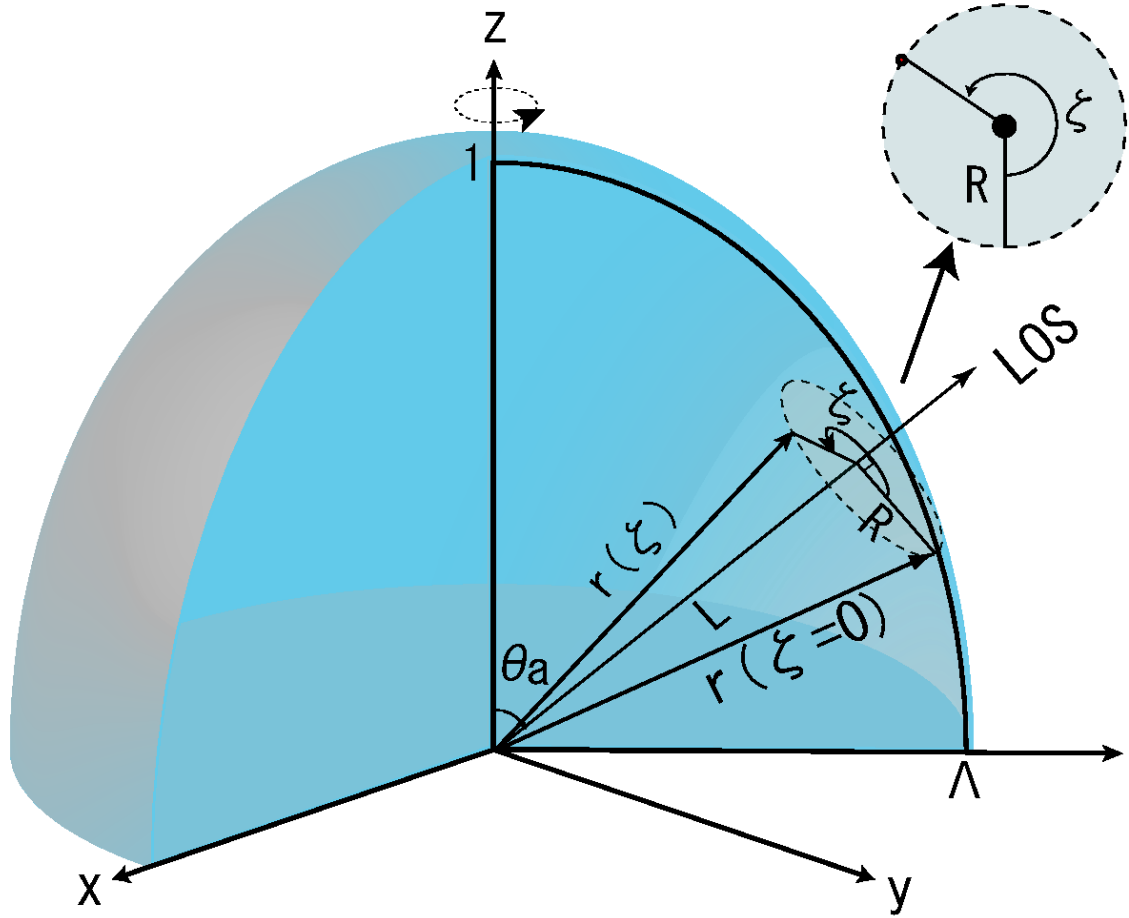}} \caption{Schematic representation of a prolate
cluster with an axis ratio $\lambda_a : \lambda_b : \lambda_c = 1 :
\Lambda : \Lambda$.  Given the symmetry around the $z$-axis (the
long axis), an LOS through the cluster center is specified by an
angle $\thetaa$ from the $z$-axis. An arbitrary position in the cluster
$\rv$ is expressed in terms of $L$ (the projection of $\rv$ onto the LOS
direction), $R$ (the projection of $\rv$ onto the plane normal to the
LOS), and $\zeta$ (the azimuthal angle on the plane normal to the
LOS). } \label{fig:sp}
\end{figure}

In this Appendix, we derive the distribution of $\fH$ due to
the asphericity of clusters, by considering the following two extreme
cases; the prolate ($\lambda_a = \lambda_b < \lambda_c$) and the oblate ($\lambda_a < \lambda_b = \lambda_c$)  ellipsoids.  We choose $z$-axis as
the long (short) axis and $x$- and $y$-axes as the short (long) axes for
an prolate (oblate) ellipsoid.  The direction of the unit vector along
the LOS of an observer, $\av$, is defined in terms of the spherical
coordinate ($\thetaa$, $\phia$).  Figure \ref{fig:sp} shows a schematic
picture of a prolate ellipsoid.

Let us define the quantity $\Lambda \equiv
\lambda_a/\lambda_c = \lambda_b/\lambda_c$ ($\Lambda \equiv
\lambda_b/\lambda_a=\lambda_c/\lambda_a$)  for the prolate (oblate)
ellipsoid.  We assume that the gas density follows the {\it prolate (oblate)} $\beta$ model:
\begin{eqnarray}
 \label{eq:opbeta}
 n(\rv)_{|\thetaa} &=& n_0 
   \left( 1 + (\rvd/\rc)^2 \right)^{-3 \beta /2}, \\
 \rvd &\equiv& |\rv| [\sin^2{\thetar}/\Lambda^2  + \cos^2{\thetar}]^{1/2} ,
\end{eqnarray}
and $\thetar$ is the angle between $z$-axis and $\rv$.  Because the
surface brightness profile is independent of $\phia$ due to the
$z$-axial symmetry, one can express $\fH$ as function of $\thetaa$.

For an isothermal cluster, the surface brightness averaged over a circle
of radius $R$ is proportional to $\int n^2 dL$ averaged over the angle
$\zeta$ of the circle. We put $\zeta=0$ where $\rv$ is located on the
same plane defined by the LOS and $z$-axis. To compute the averaged
surface brightness, we need an expression of the density $n$ as a
function of $L$, $R$, and $\zeta$.  In the Cartesian coordinate,
$\rv(\zeta=0)$ is
\begin{eqnarray}
 \rv(\zeta=0) = 
\left(
 \begin{array}{rl}
\sqrt{R^2+L^2} \cos{\phia} \sin{(\arctan{R/L}+\thetaa)} \\
\sqrt{R^2+L^2} \sin{\phia} \sin{(\arctan{R/L}+\thetaa)} \\
\sqrt{R^2+L^2} \cos{(\arctan{R/L}+\thetaa)} 
 \end{array}
\right).
\end{eqnarray}
Multiplying the rotation matrix around $\av$, $\Mr$ to $\rv(\zeta=0)$,
we obtain 
\begin{eqnarray}
 \rv(\zeta) &=& \Mr \rv(\zeta=0) \cr
&=& \left(
 \begin{array}{rl}
R \cos{\zeta} \cos{\phia} \cos{\thetaa} + R \sin{\zeta} \sin{\phia} + L
 \cos{\phia} \sin{\thetaa} \\
R \cos{\zeta} \sin{\phia} \cos{\thetaa} - R \sin{\zeta} \cos{\phia} + L
 \sin{\phia} \sin{\thetaa} \\
L \cos{\thetaa} - R \cos{\zeta} \sin{\thetaa}
 \end{array}
\right),
\end{eqnarray}
where
\begin{eqnarray}
\hspace*{-1.5cm} &&\Mr \equiv \cr
\hspace*{-1.5cm} &&\left(
\begin{array}{ccc}
\xa^2 + (1-\xa^2) \cos{\zeta} & \xa \ya (1 - \cos{\zeta}) + \za
 \sin{\zeta} & \za \xa (1 -\cos{\zeta}) - \ya \sin{\zeta} \\ 
\xa \ya (1-\cos{\zeta}) - \za \sin{\zeta} & \ya^2 + (1-\ya^2)
 \cos{\zeta} & \ya \za (1-\cos{\zeta}) + \xa \sin{\zeta} \\
\za \xa (1-\cos{\zeta})+\ya \sin{\zeta} & \ya \za (1-\cos{\zeta}) - \xa
 \sin{\zeta} & \za^2 + (1-\za^2) \cos{\zeta} 
\end{array}
\right) ,
\end{eqnarray}
and
\begin{eqnarray}
\left(
\begin{array}{rl}
\xa \\
\ya \\
\za 
\end{array}
\right)
\equiv
\left(
\begin{array}{rl}
\cos{\phia} \sin{\thetaa}\\
\sin{\phia}\sin{\thetaa}\\
\cos{\thetaa}
\end{array}
\right) .
\end{eqnarray}
Thus, we obtain
\begin{eqnarray}
 |\rv(\zeta)| \cos{\thetar} = L \cos{\thetaa} - R \cos{\zeta} \sin{\thetaa}. 
\end{eqnarray}
Then, $|\rv(\zeta)|$ and $\thetar$ are written as
\begin{eqnarray}
|\rv(\zeta)| &=& \sqrt{L^2 + R^2} \\
 \thetar &=& 
\arccos{\left(\frac{L \cos{\thetaa} - R \cos{\zeta} \sin{\thetaa}}
{\sqrt{L^2 + R^2}}\right)}. 
\end{eqnarray}
Combining with equation (\ref{eq:opbeta}),
we can write $n(\rv)_{|\thetaa}$  in terms of $L$,$R$, and $\zeta$ as
\begin{eqnarray}
 n(\rv)_{|\thetaa} 
= n_0 \left( 1 + (\rvd(R,L,\zeta)_{|\thetaa}/\rc)^2 \right)^{-3 \beta /2
 } \equiv n(R,L,\zeta),
\end{eqnarray}
where
\begin{eqnarray}
\rvd(R,L,\zeta)_{|\thetaa} 
\equiv \rvd = \frac{\sqrt{L^2 + R^2 + [\Lambda^2 - 1](L \cos{\thetaa} - R
 \cos{\zeta} \sin{\thetaa})^2 }}{\Lambda} .
\end{eqnarray}
Then, the averaged surface brightness at $R$ is 
\begin{eqnarray}
\Sx (R)_{|\thetaa} 
&=& \frac{1}{2 \pi} \int_0^{2 \pi} d \zeta \int_{-\infty}^{\infty}
  d L [n(R,L,\zeta)_{|\thetaa}]^2 \cr
&=& \frac{n_0^2 \rc}{2 \pi} \int_0^{2 \pi} d \zeta \int_{-\infty}^{\infty}
  d q_L \left[ \frac{q_L^2 + q_R^2 + (\Lambda^2 -1)(q_L \cos{\thetaa} -
  q_R \cos{\zeta} \sin{\thetaa})^2}{\Lambda^2} + 1 \right]^{-3 \beta} \cr
&\equiv& \frac{n_0^2 \rc}{2 \pi} I(q_R)_{|\thetaa} 
\end{eqnarray}
where we define the normalized length by $\rc$, $q_R \equiv R/\rc$,$q_L
\equiv L/\rc$. We compute $I(q_R)_{|\thetaa}$ numerically for
$\Lambda=0.5$ (prolate) and $\Lambda=2.0$ (oblate) adopting
$\beta=0.65$. We fit $I (q_R)_{|\thetaa}$ from $q_R=0$ to $q_R=10.0$ by
with a functional form of the surface brightness profile assuming the
spherical beta model $(\propto [1 + (q_R/
q_{c,\mathrm{fit}|\thetaa})^2]^{-3 \beta_{\mathrm{fit}|\thetaa}+1/2})$.
Thus, we obtain the counter part of $\rcisobeta$ ,
$q_{c,\mathrm{fit}|\thetaa}\equiv \rfitsx/\rc$ and the
fitted value of $\beta$, $\beta_{\mathrm{fit}|\thetaa}$.  While,
$q_{\rm c,iso\beta|\thetaa} \equiv \rcisobeta/\rc$ is written as
\begin{eqnarray}
\label{eq:rcisobetathetaa}
q_{\rm c,iso\beta|\thetaa}=(\sin^2{\thetaa}/\Lambda^2+\cos^2{\thetaa})^{-1/2}
 \frac{G(\beta) G(\beta_{\mathrm{fit}|\thetaa}/2)^2}{G(\beta/2)^2
 G(\beta_{\mathrm{fit}|\thetaa})}.
\end{eqnarray}
The first term of the right-hand side represents the elongation of the
radius toward the LOS. The second term is the correction to the use of
$\beta_{\mathrm{fit}|\thetaa}$ in observation instead of the true 
$\beta$. However, the correction is very small (within 0.01\% error).

Finally, we obtain the bias of $H_0$ as a function of $\thetaa$, 
\begin{eqnarray}
\label{eq:foftheta}
\fH (\thetaa) \equiv
 \frac{q_\mathrm{c,\mathrm{fit}|\thetaa}}{q_{\rm c,iso\beta|\thetaa}}. 
\end{eqnarray}

The probability of $\fH$ for the random assignment is proportional to
the solid angle $\Omega (\fH)$. If $\fH (\thetaa)$ is a monotonic
function, the PDF of $\fH$ is obtained as
\begin{eqnarray}
\label{eq:PfH}
 P(\fH) = \frac{1}{4 \pi}\frac{d \Omega}{d \fH} 
&=& \frac{1}{4 \pi}\frac{d \Omega}{d \thetaa}
  \Big|\frac{d \thetaa}{d \fH }\Big| \cr
 &=& \frac{\sin{\thetaa(\fH)}}{2} \Big|\frac{d \thetaa(\fH)}{d \fH}\Big|,
\end{eqnarray}
where $\thetaa(\fH) = f^{-1}_{\H} (\thetaa)$.

\begin{figure}[htbp]
{\plotone{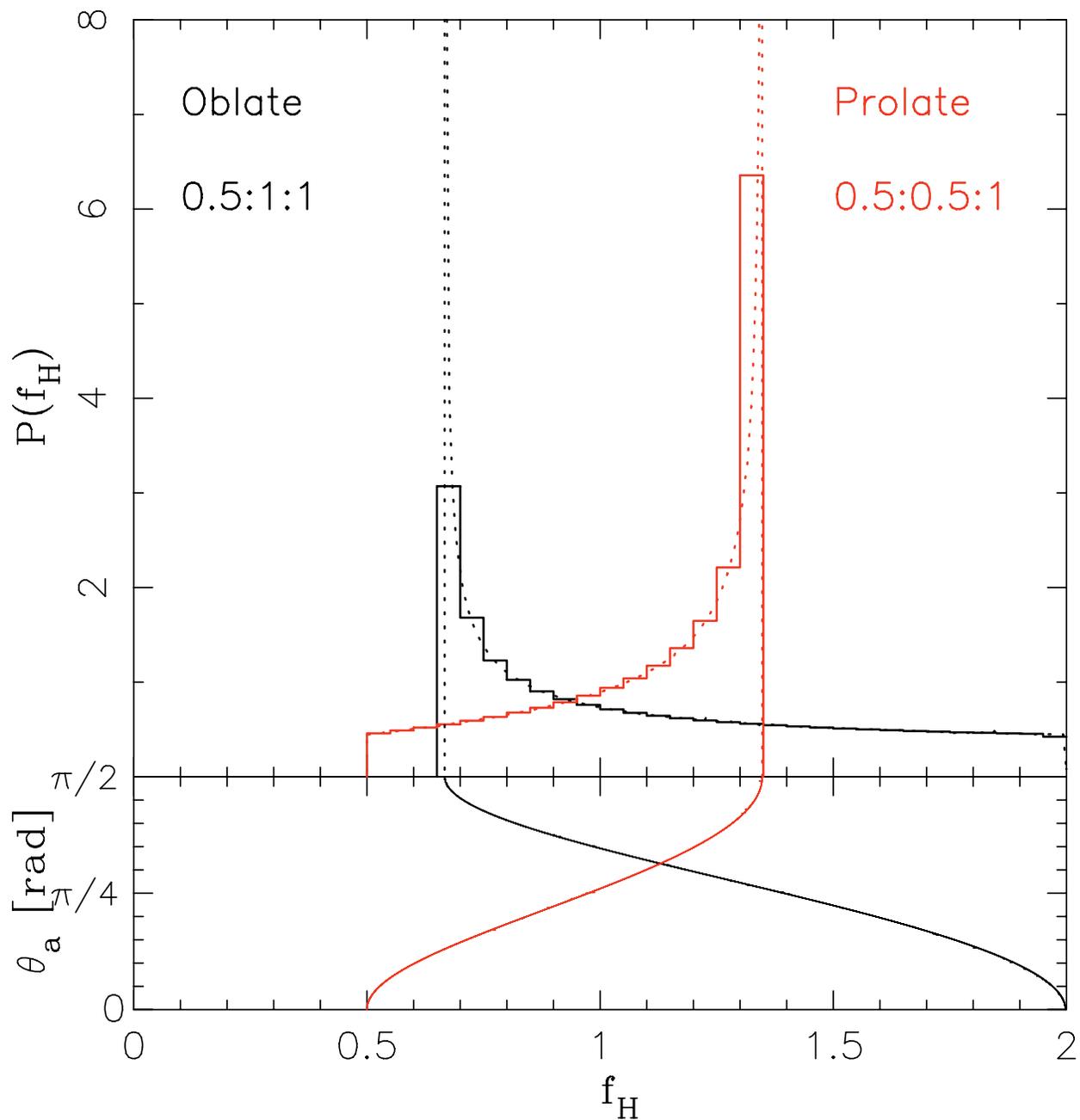}} 
\caption{Upper: The PDF of $\fH$ for the oblate ($\Lambda=2$; black curves) 
and prolate ($\lambda=0.5$; red curves) ellipsoids. 
Dotted lines represent equation (\ref{eq:PfH}),
 while solid lines show the corresponding histograms 
with a bin size of $\Delta \fH = 0.05$. Lower: The angle, $\thetaa$, as
 a function of $\fH$.}
 \label{fig:pop}
\end{figure}

Dotted lines in the upper panel of Figure \ref{fig:pop} show equation
(\ref{eq:PfH}) for prolate ($\Lambda=0.5$) and oblate ($\Lambda=2.0$)  
ellipsoids.  As shown in the lower panel, the corresponding $\thetaa$ is
a monotonically increasing (decreasing) function of $\fH$ for the prolate
(oblate) ellipsoid.  At $\thetaa=0$, $\fH$ is equal to $\Lambda$, which
corresponds to the case that the LOS is along the $z$-axis.

The PDF diverges at $\thetaa=\pi/2$. This can be
understood as follows. Equations (\ref{eq:rcisobetathetaa}) to
(\ref{eq:PfH}) imply that
\begin{eqnarray}
 P(\fH) \propto
\sin{\thetaa(\fH)} 
\Big|\frac{d q_{\rm c,iso\beta|\thetaa}^{-1}}{d \thetaa(\fH)}\Big|^{-1}
\propto  \frac{\sqrt{\cos^2\thetaa(\fH)+
\Lambda^{-2}\sin^2\thetaa(\fH)}}{\cos\thetaa(\fH)},
\end{eqnarray}
where we ignore the $\thetaa$-dependence of $q_{c,\mathrm{fit|\thetaa}}$
and $\beta_{\mathrm{fit}|\thetaa}$. Thus $\thetaa\approx \pi/2$,
$P(\fH)$ diverges as $1/\cos\thetaa$. Note, however, its integration
over a finite size of $\fH$ does not diverge (see
eq.[\ref{eq:PfH}]). This is plotted in the solid histograms, where the
bin size $\Delta\fH= 0.05$ is adopted.  The resulting distribution is
skewed positively (negatively) for the prolate (oblate) ellipsoid, which
is consistent with the results shown in Figure \ref{fig:dist}.

\bigskip
\bigskip


\end{document}